\documentclass{article}
 \usepackage{epsfig}
 \usepackage{amssymb}

\begin{document}

 \begin{flushleft}
  {KEK--TH--759\\
   MSUHEP--010401}\\   
 \end{flushleft}

 \vspace{5mm}

 \begin{center}

 {\huge
  Can the charged Higgs boson be discovered in $e\gamma$ collisions
  using the channel $e^-\gamma\to\nu H^-$?}

 \vspace{5mm}

 {Shinya Kanemura\\
  Physics and Astronomy Department, Michigan State University\\
  East Lansing, MI 48824--1116, USA}

 \vspace{3mm}

 {Kosuke Odagiri\\
  Theory Group, KEK, 1--1 Oho, Tsukuba, Ibaraki 305--0801, Japan}

 \end{center}

 {\bf Abstract:} 

 We consider the single production of charged Higgs bosons $H^\pm$ of the
two-Higgs-doublet model in $e\gamma$ collisions through the production
subprocess $e^-\gamma\to\nu H^-$.
 As the production rate is governed by the size of the loop-induced $H^\pm
W^\mp\gamma$ coupling, the cross section can only be substantial for
smaller values of $\tan\beta$, $\tan\beta\lesssim1$, where the top-bottom
loop contribution is enhanced.
 In this case, however, the natural width of the charged Higgs boson
becomes large so that the Standard Model continuum background becomes
important. We study the background subprocess $e^-\gamma\to\nu\bar t b$
including the interference with the signal, and find that in the region of
charged Higgs mass that can be interesting, the signal is difficult to
detect.

 \vspace{1cm}

 Multi-doublet Higgs sectors arise naturally in several extensions of the
Standard Model (SM) including the Minimal Supersymmetric Standard Model
(MSSM), and these give rise to the charged Higgs boson $H^\pm$.
 Its discovery at present and future collider experiments has been the
subject of much debate\footnote{See ref.~\cite{kmo_epem} and the
references therein for a review of the current status.}, but it remains
that the large mass region is elusive.

 Here we study the single production channel,
 \begin{equation}e^-\gamma\to \nu H^-,\label{signal}\end{equation}
 in the $e\gamma$ option of future linear colliders.
 \begin{figure}[ht]
 \centerline{\epsfig{file=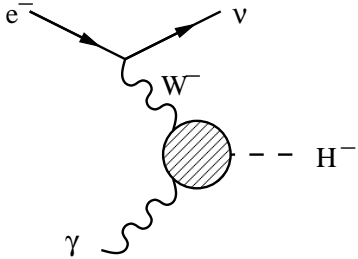}}
 \caption{The Feynman graph for the signal process}
 \label{feyn}
 \end{figure}
 The process involves the $H^\pm W^\mp\gamma$ vertex \cite{peyranere}
which is forbidden at tree level because of gauge invariance, and is 
therefore loop-induced.
 The main motivation for considering this process is as a discovery mode.
 Although the physics potential of the $e\gamma$ option is limited
compared to the $e^+e^-$ and $\gamma\gamma$ options, it is possible that
earlier experiments find one or more of the neutral Higgs bosons and the
measurements indicate the presence of a heavy charged Higgs boson in the
parameter region that is accessible using the above channel. For this
reason it is important to have a grasp of the accessibility of the
$e^-\gamma\to\nu H^-$ channel and the parameter regions in which it is
applicable.

 To be more specific, we want to find out whether a charged Higgs boson
heavier than a half of the $e^+e^-$ centre-of-mass energy can be produced
in the $e^-\gamma$ environment for values of $\tan\beta$ and other
parameters which are inaccessible through the $e^+e^-$ single charged
Higgs production modes studied in ref.~\cite{kmo_epem}.

 For evaluating the signal cross section we utilised the form factors as
given in ref.~\cite{shinya_form}.
 Out of the three form factors describing the $H^\pm W^\mp_\mu\gamma_\nu$
vertex, only two are independent by gauge invariance such that the vertex
is written as:
 \begin{equation}
 V^{\mu\nu} =
 G\frac{-p_\gamma\cdot p_Wg^{\mu\nu}+p^\mu_\gamma p^\nu_W}{m_W^2}
 +iH\frac{p^\rho_\gamma p^\sigma_W{\varepsilon^{\mu\nu}}_{\rho\sigma}}{m_W^2}.
 \label{formfactors}
 \end{equation}
 Our convention is $\varepsilon_{0123}=1$. Our spin averaged matrix
element squared, corresponding to the Feynman diagram of figure
\ref{feyn}, is given by:
 \begin{equation}
 \overline{|{\mathcal M}|^2}=
 \left(\frac{e^2\sqrt{-\hat{t}}}{4m_W\sin^2\theta_W(\hat{t}-m_W^2)}\right)^2
 \left[\hat{s}^2|G-H|^2+\hat{u}^2|G+H|^2\right].
 \end{equation}
where $\hat{s}, \hat{t}$ and $\hat{u}$ are the usual Mandelstam variables 
for the subprocess. 

 The form factors $G$ and $H$ were calculated in the two-Higgs-doublet
model (2HDM) with softly broken discrete symmetry. For the Yukawa
interaction of quarks, we have adopted the so-called Type II coupling
\cite{typeII}. For the fermionic loops, we only include the top-bottom
contributions.

 \begin{figure}[ht]
 \centerline{\epsfig{file=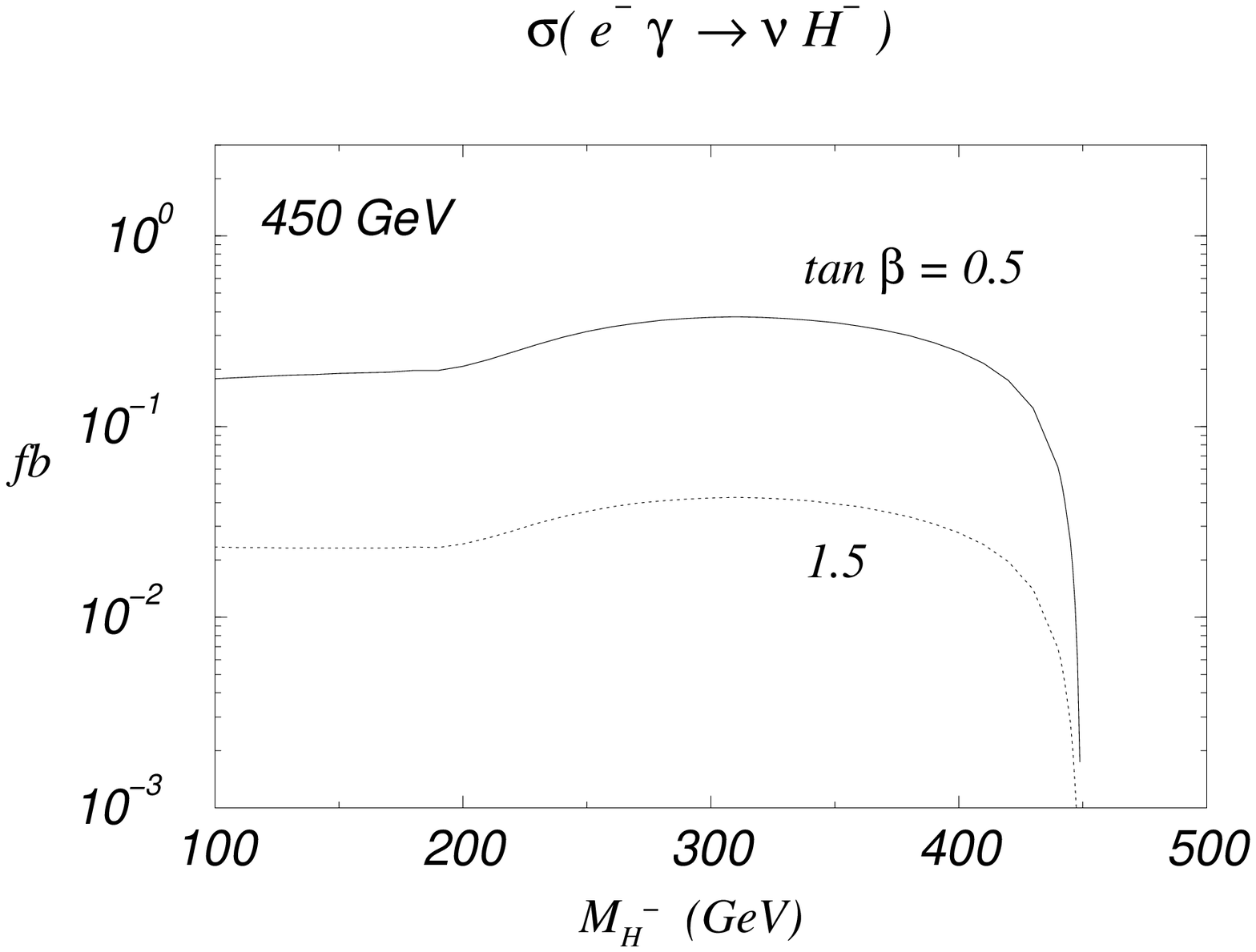,width=6.2cm,angle=0}
             \epsfig{file=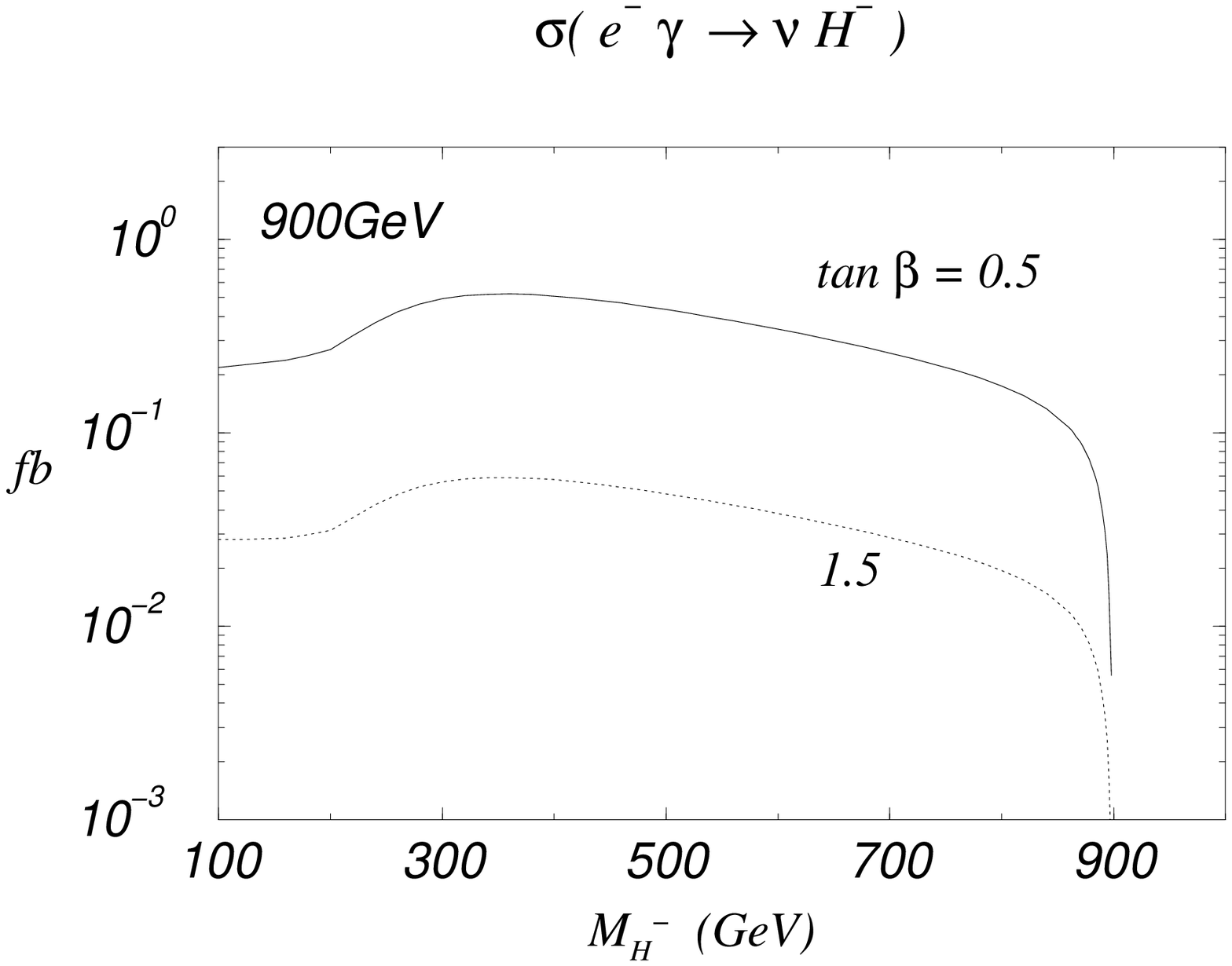,width=6.2cm,angle=0}}
 \caption{The signal cross section at $\sqrt{\hat s}=450$ GeV (left) and
          $\sqrt{\hat s}=900$ GeV (right).}
 \label{rate_signal}
 \end{figure}

 For the Standard Model parameters, we followed the values in
ref.~\cite{kmo_epem}. For the parameters of the 2HDM Higgs sector, we
chose
 $\alpha=\beta-\frac\pi2$, 
 $m_{h^0}=120$ GeV, $m_{H^0}=\sqrt{m_{H^\pm}^2+m_W^2}$, 
 $m_{A^0}=\sqrt{m_{H^\pm}^2-m_W^2}$, and $\mu=m_{A^0}$.
 $\alpha$ is the mixing angle between CP-even neutral Higgs bosons.
 $m_{h^0}$, $m_{H^0}$ and $m_{A^0}$ are the masses of the lighter and
heavier CP-even Higgs bosons $h^0,H^0$ and the CP-odd Higgs boson
$A^0$, respectively.
 $\mu$ is the soft-breaking mass parameter for the discrete symmetry. This
parameter choice corresponds to the MSSM Higgs sector in the large
$m_{H^\pm}$ limit.

 Figure \ref{rate_signal} shows the total signal cross section at two
values of subprocess centre-of-mass energy $\sqrt{\hat{s}}$, 450 GeV and
900 GeV. These roughly correspond to $e\gamma$ collisions using Compton
back-scattered photons from a $e^+e^-$ collider at $\sqrt{s}=500$ GeV and
1 TeV. We have omitted the photon structure function for simplicity and in
order that the physics of the process becomes more clear. The effect of
this convolution with the photon structure function is that the large mass
end of the distribution is suppressed, and the rise of the cross section
in the medium mass range is counteracted, such that the distributions
shown in figure \ref{rate_signal} become more `flat' as seen in figure 8
of ref.~\cite{kmo_egamma}.

 We show the rates for two values of $\tan\beta$ at 0.5 and 1.5.
 As the dominant part of the cross section comes from the top-bottom
loops, the $\tan\beta$ dependence is mainly due to the Yukawa coupling and
the chirality structure of the $H^\pm W^\mp\gamma$ vertex. In the large
and small $\tan\beta$ regions, the $\tan\beta$ dependence is approximately
as follows:
 \begin{eqnarray}
    \sim m_W^4 \cot^2 \beta \;\;
    \left(\tan\beta \ll \frac{m_t}{m_b}\right),
 \;\; {\rm and}   \;\; \sim m_W^4 \frac{m_b^4}{m_t^4} \tan^2 \beta
             \;\;  \left(\tan\beta \gg \frac{m_t}{m_b}\right).
 \end{eqnarray}
 The enhancement of the cross section in the large $\tan\beta$ region is
negligible. The Higgs and gauge boson contributions are small when the
cross section is substantial. We have explicitly confirmed that other
combinations of Higgs parameters do not affect the rate significantly.  
This is a reflection of the fact that there are no ${\cal O}(M^2/m_W^2)$
terms in the $H^\pm W^\mp\gamma$ vertex due to gauge invariance, and only
${\cal O}(\ln M^2)$ terms contribute for large $M$. $M$ represents the
characteristic mass of the particles in the loop.

 From this $\tan\beta$ dependence, it follows that if we adopt the MSSM
parameters based on the LEP constraints \cite{LEP_MSSM} which give
$\tan\beta\gtrsim3$, the signal is too small to be observable. The
$e\gamma$ integrated luminosity is typically $\sim 100$ fb$^{-1}$
\cite{TESLA}.

 The cross section rises slowly with increasing $\sqrt{\hat{s}}$ as
$\ln(\hat{s}/m_W^2)$, as this is a $t$-channel process. Our results are
substantially smaller than the case of neutral Higgs boson production
considered in ref.~\cite{repko}. The case of neutral Higgs boson
production is dominated by the low $p_T$ photon-fusion contributions, such
that the cross section scales as $\ln(\hat{s}/m_e^2)$.

\begin{figure}[ht]
 \centerline{\epsfig{file=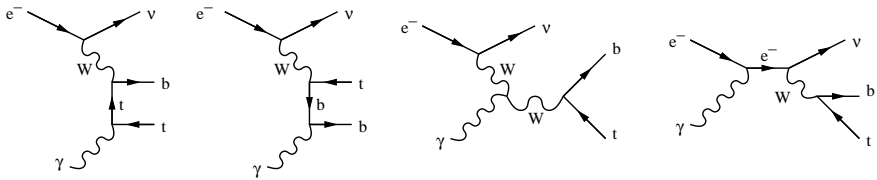,width=12cm}}
 \caption{The Feynman graphs for the background process}
 \label{bkgd}
 \end{figure}

 In the heavy $H^\pm$ region which is the main interest of our study, the
branching ratio for the mode $H^-\to\bar tb$ is practically 100\%
especially at small $\tan\beta$. We do not consider other decay modes as
they offer no advantage compared to this mode. For the production process
(\ref{signal}) and the $\bar tb$ decay mode, there is irreducible
background coming from the Standard Model continuum production:
 \begin{equation}e^-\gamma\to \nu\bar tb.\label{background}\end{equation}
 We evaluated this background, as well as its interference with the signal
which we discuss later on, using HELAS \cite{helas}. The Feynman diagrams
for this process are shown in figure \ref{bkgd}. The numerical
integrations were carried out using a combination of Simpson's rule and
naive Monte Carlo.
 We set all widths to zero, except for the natural width of the charged
Higgs boson $\Gamma_{H^\pm}$ which affects the calculation of the signal
and the signal--background interference.

\begin{figure}[ht]
 \centerline{\epsfig{file=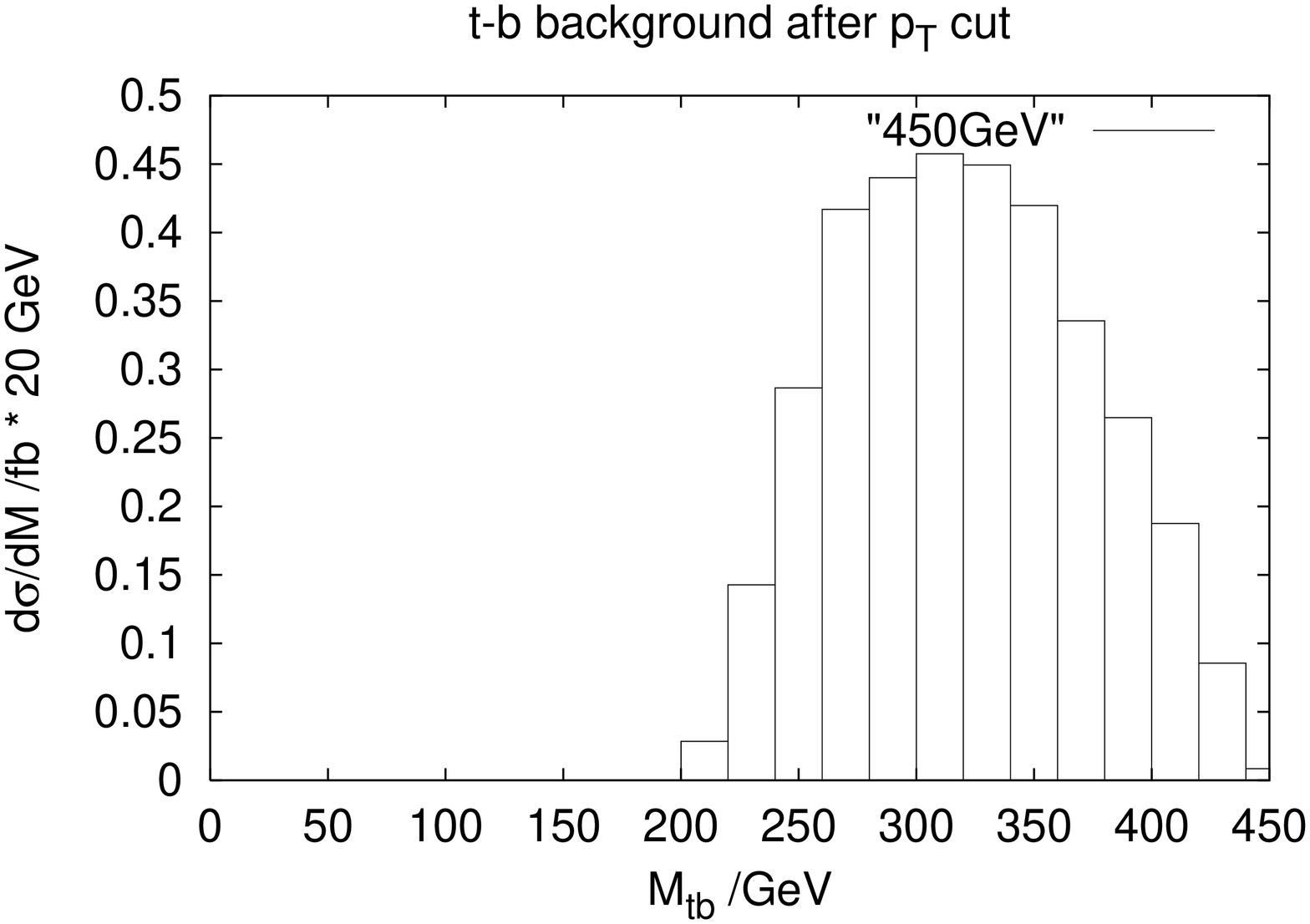,width=4.5cm,angle=270}
             \epsfig{file=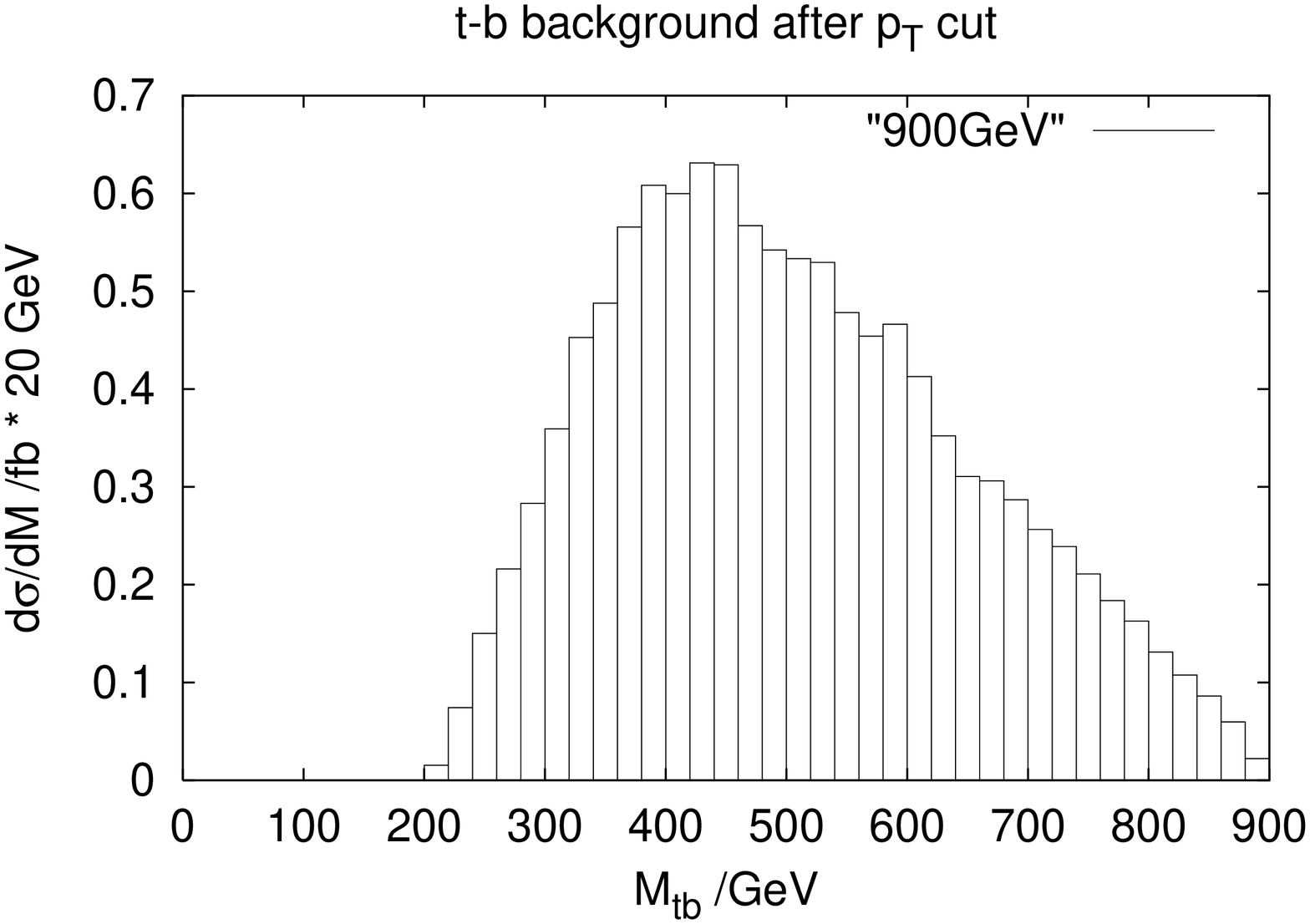,width=4.5cm,angle=270}}
 \caption{The background cross section at $\sqrt{\hat s}=450$ GeV (left)
          and $\sqrt{\hat s}=900$ GeV (right).}
 \label{rate_bkgd}
 \end{figure}

 The total rate for the background at $\sqrt{\hat s}$ of 450 GeV and 900
GeV are 28.4 fb and 69.2 fb, respectively. However, these numbers are not
very meaningful as the dominant part of the background cross section comes
from regions where the bottom quark is collinear with the initial photon
direction. In order to have a more meaningful comparison with the signal
cross section we introduced a cut on the bottom quark $p_T$ at 50 GeV and
100 GeV respectively for the two collider energies. We have explicitly
verified that our main conclusions are independent of the value of this
cut-off.
 The background cross sections drop to 3.5 fb and 11.8 fb, respectively,
and the resulting $\bar tb$ invariant mass distributions are shown in
figure \ref{rate_bkgd}. The binning width is 20 GeV.

 20 GeV is a pessimistic estimate for the mass reconstruction resolution
so that, had the widths been dominated by the detector resolution, it
would be meaningful to compare the numbers from figure \ref{rate_signal}
directly with those on figure \ref{rate_bkgd}.
 One may thereupon conclude for instance that for $\tan\beta=0.5$,
$m_{H^\pm}=700$ GeV and $\sqrt{\hat{s}}=900$ GeV, the signal and
background rates are both about 0.3 fb so that for an integrated
luminosity of 100 fb$^{-1}$ and an optimistic value for the acceptance
rate of about 25\% let us say, there would be a nearly 3 $\sigma$ signal.

 However, this is not the case as when $\tan\beta$ is small and the signal
is large, the natural width $\Gamma_{H^\pm}$ also becomes large. For the
case $\tan\beta=0.5$ and $m_{H^\pm}=700$ GeV, the width due to the $\bar
tb$ decay is about 150 GeV. What happens here is that the signal and the
decay width both behave as $\cot^2\beta$ at low $\tan\beta$ such that the
signal distribution in $M_{\bar tb}$ near $m_{H^\pm}$ remains constant
when $\tan\beta$ is varied. Hence S/B, the signal over the continuum
background integrated over the resonant region, remains constant as a
function of $\tan\beta$.
 However, the total number of signal events does increase at low
$\tan\beta$ such that the signal significance, defined by S/$\sqrt{{\rm
B}}$ over the resonant region, does improve at low $\tan\beta$.


 Let us adopt the naive approach, which we nevertheless believe gives a
good estimate of the full result, in which we define the cross section
over the resonant region by:
 \begin{equation}\label{defn_NWA}
  \sigma_{\rm{resonance}}=\frac{d\sigma}{dM_{\bar tb}}
  \Bigr|_{M_{\bar tb}=m_{H^\pm}}\times \Gamma_{H^\pm}.
 \end{equation}
 We also include the effect of the interference between the signal and the
background in order to have a consistent description of the production
process.
 This definition circumvents the complications associated with the 
treatment of the signal when the charged Higgs boson is significantly 
off-shell.
 For a signal that has a pure Breit-Wigner distribution, (\ref{defn_NWA})  
gives a rate that is too small by factor $\pi/2$. Our argument is that we
are only interested in the resonant region whose width is given by
$\Gamma_{H^\pm}$. We note that this approach gives a somewhat optimistic
estimate, where it is assumed that both the signal and the background are
nearly flat over the resonant region whereas the signal is peaked when
$H^\pm$ is on-shell.
 We also note that this naive approach becomes questionable when the width
becomes comparable with the characteristic mass scale. However, our goal
in this study is merely to establish whether, and in what region of the
phase space, the signal process (\ref{signal}) can be seen.

 S/B remains almost constant as a function of $\tan\beta$ as mentioned
before, so that we can do this calculation for any small $\tan\beta$. We
adopt the value $\tan\beta=0.5$, and in table \ref{interference} we show
the total rate against the SM expectation over the resonant region as
defined above. This value of $\tan\beta$ is near the lowest bound
acceptable from the criterion of the validity of perturbation theory
\cite{unitarity}.

 \begin{table}[ht]\begin{center}
 \begin{tabular}{|c|c|c|c|}\hline
 $M_{H^\pm}$ /GeV & $\Gamma_{H^\pm}$ /GeV & total /fb & background /fb\\\hline
  250 & 15.7 & 0.36 & 0.23\\
  300 & 31.6 & 0.94 & 0.75\\
  350 & 47.6 & 1.13 & 0.97\\
  400 & 63.2 & 0.78 & 0.70\\\hline
 \end{tabular}\\[2mm]{$\sqrt{\hat s}=450$ GeV}\\[5mm]
 \begin{tabular}{|c|c|c|c|}\hline
 $M_{H^\pm}$ /GeV & $\Gamma_{H^\pm}$ /GeV & total /fb & background /fb\\\hline
  300 & 31.6 & 0.66 & 0.51\\
  400 & 63.2 & 2.17 & 2.05\\
  500 & 93.0 & 2.90 & 2.87\\
  600 & 121. & 2.45 & 2.52\\
  700 & 149. & 2.07 & 2.14\\
  800 & 175. & 1.19 & 1.25\\\hline
 \end{tabular}\\[2mm]{$\sqrt{\hat s}=900$ GeV}
 \end{center}
 \caption{Total rate versus the expected background rate over the resonant
region, as defined in the text, at two centre-of-mass energies, after
the $p_T$ cut. The numbers shown are for $\tan\beta=0.5$.}
 \label{interference}\end{table}

 From the numbers shown in table \ref{interference}, we point out the
following. First, let us consider the $e\gamma$ integrated luminosity of
100 fb$^{-1}$ per year and an acceptance rate of few times 10\%. For
concreteness let us adopt 25\% for example as an optimistic estimate. We
see that the signal significance, defined as S/$\sqrt{{\mathrm B}}$ with S
being defined here as the total rate minus the background rate, is at most
about $1.25\sigma$ at $\sqrt{\hat{s}}=450$ GeV and $1\sigma$ at
$\sqrt{\hat{s}}=900$ GeV. Even with increased acceptance rate and
increased luminosity, the channel is not useful for discovery in this 
model or generally in the 2HDM.

 Second, the interference between the signal and the background is
negative and sometimes large. In fact, there are regions where the total
rate is smaller than the expected background. This is explained as
follows. According to the definition of equation (\ref{defn_NWA}), it is
easy to see that the interference between the signal and the background is
due to the imaginary part of the form factors $G$ and $H$ given in
equation (\ref{formfactors}). The imaginary parts of the top-bottom loop
contributions in these form factors have exactly the form that comes from
the interference between the continuum background and the decay
$H^-\to\bar tb$. Thus it is possible to relate the magnitude of the
interference term to the signal, and it turns out that the contribution of
the interference term is of the same order and has the sign that is
opposite to the signal.

 Lastly, we note that in this model the channel seems to offer no
advantage compared to $W^\pm H^\mp$ associated production in the $e^+e^-$
mode \cite{kmo_epem, shinya_form, eewh}.

 The extension of our calculation by the inclusion of other light fermions
into the loop can not improve the situation as whatever increases the
signal rate also increases the charged Higgs boson width while the
branching ratio into $\bar tb$ in general falls.
 On the other hand, by the inclusion of new heavy virtual particles in the
$H^\pm W^\mp\gamma$ vertex that have large non-decoupling contributions,
it is possible that the signal cross section is enhanced significantly
without enhancing the width.
 This may be possible if we consider non-decoupling contributions from
heavy squarks with large left--right mixings in the stop sector, and in
this case, there is some discovery potential for this mode.

 In this paper we did not discuss the decay of the top quark. 
 The polarisation of the top quark is opposite between the signal and 
the background in the limit $M_{H^\pm}\gg m_t$, but the statistics 
is presumably too low to utilise polarisation analysis as described 
in ref.~\cite{top_polarisation}. 
 We should also mention that in our study we have neglected the 
reducible background from the top pair production subprocess 
$e^-\gamma\to e^-t\bar t$. This could very well be important.
 The consideration of these more detailed points is expected to 
make signal detection even more difficult.

 To conclude, we have considered the process $e^-\gamma\to\nu H^-$ as a
discovery mode for the charged Higgs boson at the $e\gamma$ option of
future linear colliders.
 Although the signal evaluated in the 2HDM can reach reasonable rates for
$\tan\beta\lesssim1$, the background is large and it is difficult to see
the charged Higgs boson through this channel.

Finally, we acknowledge advice and technical help from Stefano Moretti.  
We also thank Yasuhiro Okada, Wayne Repko and C.--P. Yuan for enjoyable
discussions. We are grateful for the financial support from KEK which made
this collaboration possible.

 \end{document}